%% file: wasp166.tex
\documentclass[fleqn,usenatbib]{mnras}

\usepackage{newtxtext,newtxmath}
\usepackage[T1]{fontenc}
\usepackage{ae,aecompl}

\usepackage{graphicx}	% Including figure files
\usepackage{amsmath}	% Advanced maths commands
\usepackage{amssymb}	% Extra maths symbols

\input{commands.tex}

\input{properties.tex}

\usepackage{color}

\title[NGTS Simultaneous Observations]{Simultaneous TESS and NGTS Transit Observations of \Nplanet}

\author [E. M. Bryant]{
\parbox{\textwidth}{
Edward M.~Bryant,$^{1, 2}$\thanks{E-mail: edward.bryant@warwick.ac.uk}
Daniel~Bayliss,$^{1}$
James~McCormac,$^{1,2}$
Peter~J.~Wheatley$^{1,2}$
Jack S. Acton,$^{3}$
David~R.~Anderson,$^{1,2}$
David J. Armstrong,$^{1,2}$
Fran\c{c}ois Bouchy,$^{3}$
Claudia Belardi,$^{4}$
Matthew R. Burleigh,$^{4}$
Rosie H. Tilbrook,$^{4}$
Sarah L. Casewell,$^{4}$
Benjamin F. Cooke,$^{1,2}$
Samuel Gill,$^{1,2}$
Michael R.~Goad,$^{4}$
James S. Jenkins,$^{5,6}$
Monika Lendl,$^{3,7}$
 Don~Pollacco,$^{1,2}$
 Didier Queloz,$^{9}$
Liam Raynard,$^{4}$
Alexis~M.~S.~Smith,$^{8}$
Jose~I.~Vines,$^{5}$
Richard~G.~West,$^{1,2}$
Stephane Udry$^{3}$
}
\\
$^{1}$Dept.\ of Physics, University of Warwick, Gibbet Hill Road, Coventry CV4 7AL, UK\\
$^{2}$Centre for Exoplanets and Habitability, University of Warwick, Gibbet Hill Road, Coventry CV4 7AL, UK\\
$^{3}$Observatoire de Gen{\`e}ve, Universit{\'e} de Gen{\`e}ve, 51 Ch. des Maillettes, 1290 Sauverny, Switzerland\\
$^{4}$Department of Physics and Astronomy, University of Leicester, University Road, Leicester, LE1 7RH, UK\\
$^{5}$Departamento de Astronom\'ia, Universidad de Chile, Casilla 36-D, Santiago, Chile\\
$^{6}$ Centro de Astrof\'isica y Tecnolog\'ias Afines (CATA), Casilla 36-D, Santiago, Chile\\
$^{7}$Space Research Institute, Austrian Academy of Sciences, Schmiedlstr. 6, 8010 Graz, Austria\\
$^{8}$Institute of Planetary Research, German Aerospace Center, Rutherfordstrasse 2, 12489 Berlin, Germany\\
$^{9}$Cavendish Laboratory J J Thomson Avenue Cambridge, CB3 0HE, UK
}

\date{Accepted April 15, 2020. Received April 01, 2020; in original form February 17, 2020}

\pubyear{2020}

\begin{document}
\label{firstpage}
\pagerange{\pageref{firstpage}--\pageref{lastpage}}
\maketitle

\begin{abstract}
We observed a transit of \Nplanet\ using nine NGTS telescopes simultaneously with TESS observations of the same transit.  We achieved a photometric precision of 152\,ppm per 30\,minutes with the nine NGTS telescopes combined, matching the precision reached by TESS for the transit event around this bright (T=\NshortTmag) star.  The individual NGTS light curve noise is found to be dominated by scintillation noise and appears free from any time-correlated noise or any correlation between telescope systems.  We fit the NGTS data for $T_C$ and $R_p/R_*$. We find $T_C$ to be consistent to within $0.25\sigma$ of the result from the TESS data, and the difference between the TESS and NGTS measured $R_p/R_*$ values is $0.9\sigma$. This experiment shows that multi-telescope NGTS photometry can match the precision of TESS for bright stars, and will be a valuable tool in refining the radii and ephemerides for bright TESS candidates and planets. The transit timing achieved will also enable NGTS to measure significant transit timing variations in multi-planet systems.
\end{abstract}


%%%%%%%%%%%%%%%%%%%%%%%%%%%%%%%%%%%%%%%%%%%%%%%%%%

%%%%%%%%%%%%%%%%% BODY OF PAPER %%%%%%%%%%%%%%%%%%

\section{Introduction}
\label{sec:intro}
The Transiting Exoplanet Survey Satellite \citep[TESS;][]{TESS} has been hunting for exoplanets transiting bright stars since July 2018. During its two year nominal mission, TESS will monitor $\sim 80\%$ of the sky, providing light curves at a 2\,minute cadence for $>$ 200,000 stars and at a 30\,minute cadence for the full field-of-view. TESS has already made a number of notable discoveries of exoplanets transiting bright stars, including Pi Mensae\,c \citep{huang18, gandolfi18}, HD1397\,b \citep{brahm19, nielsen19}, and the HR858 system \citep{vanderburg19}.

As part of this discovery process, ground-based transit photometry is used to improve spatial resolution, verify the transit is achromatic, check for transit timing variations, and improve the precision on transit parameters such as period, phase, and depth \citep[e.g.][]{collins18}.  However ground-based photometric observatories struggle to reach the precision of TESS for bright stars.  The primary obstacle is that precise ground-based time-series photometry requires similar magnitude reference stars to the target star in order to adequately correct for photometric changes caused by the Earth's atmosphere.  However for a T=\NshortTmag\ star the nearest similar magnitude reference star is, on average, separated by 1\,deg.  This is far out-side the field-of-view of most photometric facilities \citep{collins18}.

The Next Generation Transit Survey \citep[NGTS;][]{NGTS_2018} is an exoplanet hunting facility situated at ESO's Paranal Observatory in Chile. It consists of twelve robotic telescopes, each with a 20\,cm diameter and a field-of-view of 8\,deg$^2$. The wide field-of-view of NGTS places it in a unique position to obtain photometric follow-up from the ground of the brightest exoplanet host stars.  One limitation is that the 20\,cm telescope apertures do not collect as many photons we would wish for high-precision photometry.  We mitigate this limitation with NGTS by observing a target star simultaneously with multiple telescopes.  This increases the effective collecting area of the facility, allowing us to obtain a light curve equivalent to a larger aperture telescope while preserving the wide field-of-view.

One of the main sources of photometric noise for NGTS observations of bright stars is scintillation noise.  Scintillation noise arises from light passing through regions of turbulence in the Earth's atmosphere, resulting in changes in intensity \citep[eg.][]{osborn15}.  Scintillation has been shown to behave as white noise on transit timescales \citep{fohring19}. As such, simultaneous observations also reduce the noise present in the light curves due to atmospheric scintillation by a factor $\sqrt{N}$, where $N$ is the number of telescopes used. This is provided that scintillation noise does not correlate between the telescopes. This is especially important for bright stars ($I\leq10$) as in this regime scintillation noise dominates all other noise sources \citep{NGTS_2018}.

An early experiment using multiple NGTS telescopes was the observation of the K2 transiting planet HD106315c in 2017 \citep{smith2020}.  This work demonstrated that by using multiple NGTS telescopes we could achieve the photometric precision needed to detect a shallow (0.1\%) transit from the ground.  

In this paper we present a study into the photometric precision and noise properties of the NGTS multi-telescope observations for a bright star hosting a transiting planet.  \Nstar\ is a $V=$ \NVmag\ F-type dwarf star which hosts a \NradiusNGTS\,\rjup\ transiting exoplanet which orbits the star on a period of \Nperiodshort\ days \citep{WASP-166}.  A summary of the main properties of \Nstar\ are set out in Table~\ref{tab:stellar}. We were able to observe \Nstar\ simultaneously with TESS, which provides a unique opportunity to rigourously compare the NGTS ground-based light curve with the TESS light curve.

We set out this study in the following manner.  In Section~\ref{sec:obs} we outline both the NGTS and the TESS observations of \Nstar.  In Section~\ref{sec:noise_props} we examine the noise properties of the multi-telescope NGTS photometry. We model the NGTS data and examine the uncertainties on key system parameters in Sections~\ref{sec:NGTSfitind} \& \ref{sec:NGTScombined}. The comparison between the NGTS and TESS photometry is set out in Section~\ref{sec:ng_tess_comp}. Finally, we provide a discussion of our results in Section~\ref{sec:discussion}.

\begin{table}
	\centering
	\caption{Stellar Properties for \Nstar}
	\begin{tabular}{lcc} % six columns, alignment for each
	\hline
	\multicolumn{3}{l}{}\\
	\textbf{Stellar Parameters}	&	\textbf{Value}	&\textbf{Source}\\
	\hline
	Name        &   \Nstar      &\citet{WASP-166}\\
	TIC ID      &   \NTICID     & TIC v8\\
	TOI ID      &   \NTOIID     & \\
	\\
	R.A.		&	\NRA		&2MASS	\\
	Dec			&	\NDec		 &2MASS	\\
    $\mu_{{\rm R.A.}}$ (\masy) & \NpropRA & GAIA DR2 \\
	$\mu_{{\rm Dec.}}$ (\masy) & \NpropDec & GAIA DR2 \\
	Parallax (mas)  & \Nplx  & GAIA DR2 \\
    \rstar (\rsun) & \NstarradiusTIC      & TIC v8\\
    \mstar (\msun) & \NstarmassTIC          & TIC v8\\
    \\
	TESS        &   \NTESSmag      & TIC v8 \\
    GAIA        &   \NGAIAmag      & GAIA DR2 \\
    V           &   \NVmag         &   \\
    B           &   \NBmag         &     \\
    J           &   \NJmag         & 2MASS  \\
    H           &   \NHmag         & 2MASS  \\
    K           &   \NKmag         & 2MASS  \\
    WISE ($3.4\mu$)& \NWmag        & WISE   \\
    WISE ($4.6\mu$)& \NWWmag       & WISE   \\
    \\
    \hline
    \multicolumn{3}{l}{2MASS \citep{2MASS};  WISE \citep{WISE}; }\\
    \multicolumn{3}{l}{GAIA DR2 \citep{GAIA}; TIC v8 \citep{TESS_CAT}}\\
	\end{tabular}
    \label{tab:stellar}
\end{table}

\section{Observations}
\label{sec:obs}

\subsection{NGTS Photometry}
\Nstar\ was observed with NGTS on the nights 2019 February 25 and 26. The observations were taken at airmass$< 2$ and under photometric conditions. Across both nights, nine of the twelve NGTS telescopes were available to be used to simultaneously observe \Nstar. Each telescope used the custom NGTS filter (520 - 890\,nm). A total of 40003 images were taken across the two nights, all with an exposure time of 10\,seconds. The NGTS cameras have a readout time of approximately 3\,seconds, and therefore the full cadence of these observations was 13\,seconds. For all observations, due to the brightness of \Nstar\, the telescopes were slightly defocussed, in order to avoid saturation. The NGTS telescope guiding is performed using the DONUTS auto-guiding algorithm \citep{DONUTS}; this extremely high precision guiding resulted in a mean RMS of the location on the CCD of the target of 0.27\,pixels (plate scale: 5\,\arcsec pixel$^{-1}$) across the two nights.

The reduction is carried out on the raw frames using a custom aperture photometry pipeline which utilises SEP \citep{sextractor1996, Barbary16sep}.  This pipeline computes the photometry for a set of circular apertures with a range of radii, and the final light curve is determined by finding the aperture with the minimum RMS scatter.  Background subtraction is performed by creating and subtracting a global background map. This background map is created using SEP. After extensive testing we find that bias, dark, and flat-field frame corrections do not improve the precision of the photometry. In some conditions these corrections decrease the precision. Therefore, we do not apply these corrections during the image reduction.

Comparison stars are ranked according to their colour, brightness and position on the image, relative to the target star. This ranking is used to select the best non-saturated non-variable comparison stars. Any comparison stars with a significantly bright GAIA neighbour which would contaminate the photometric aperture are automatically discarded. The differential target star flux is computed by dividing the photometric counts from the target star by the total sum of the counts from all the comparison stars.

Despite \Nstar\ being a bright star (T=\NshortTmag), the wide field-of-view of NGTS allowed us to monitor 23 good quality comparison stars of similar magnitude to \Nstar, more than could be monitored by a 1\,m telescope. The 23 comparison stars monitored were the same for each telescope. We combine and analyse these photometric data from each NGTS telescope in Section~\ref{sec:analysis}. 

\begin{figure*}
	\includegraphics[width=\textwidth]{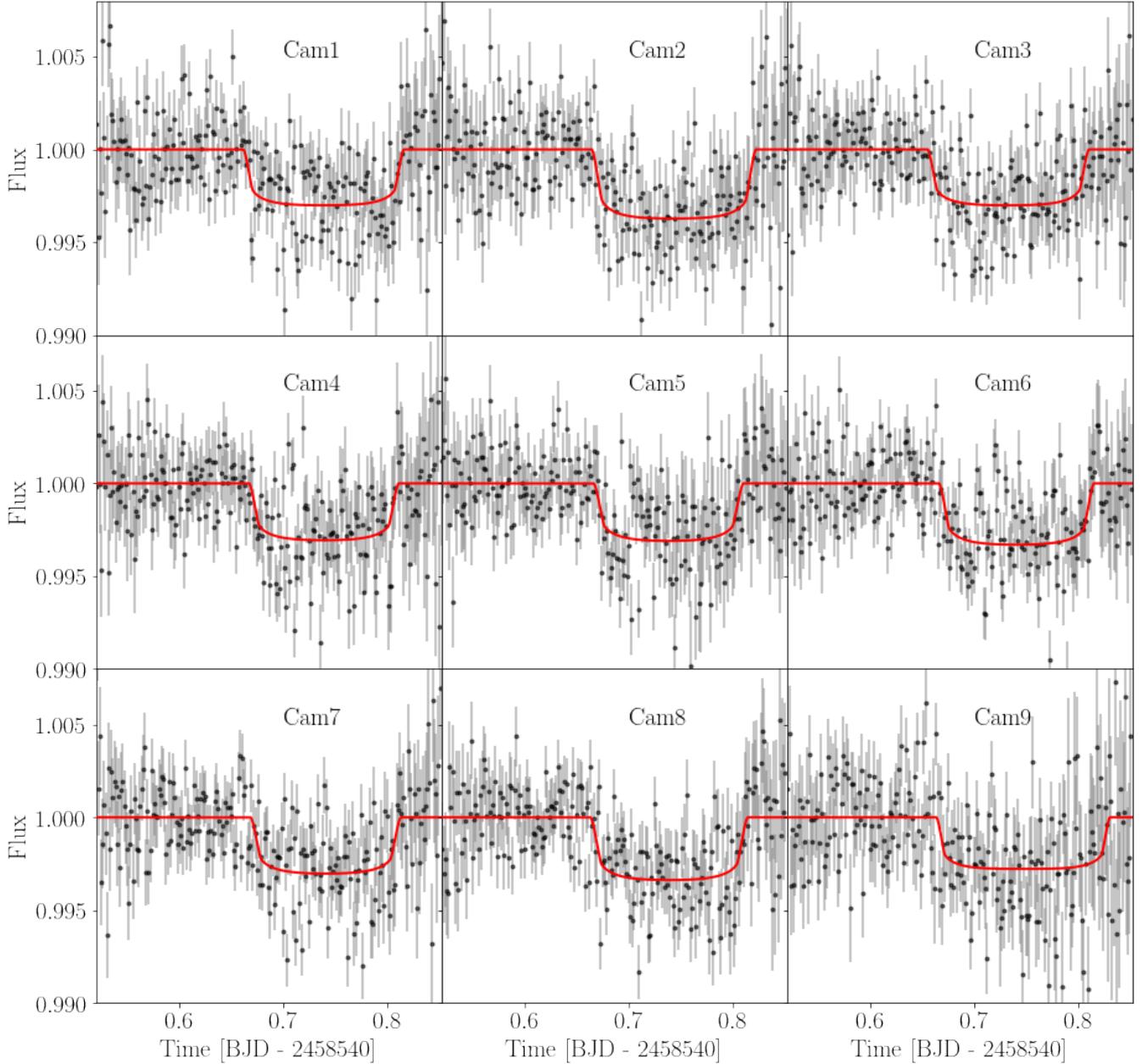}
    \caption{The nine individual NGTS telescope light curves for the transit of \Nplanet\ on 2019 February 25. The normalised flux data points are all shown binned to 2\,minutes. The solid red lines give the individual transit models for each light curves (see Section~\ref{sec:analysis} for details).}
    \label{fig:ind_lcs}
\end{figure*}

On the night of 2019 February 25, a transit of \Nplanet\ was observed, and we use these data to investigate the precision with which we can measure planetary system parameters with this observing mode. We use the out-of-transit data from the night of 2019 February 26 to investigate the multi-telescope noise properties of NGTS, as it is independent of any transit modeling steps. The nine individual NGTS telescope light curves are displayed in Figure~\ref{fig:ind_lcs}.

\subsection{TESS Photometry}
\Nstar\ was observed by TESS \citep{TESS} at a cadence of 2\,minutes from 2019 February 02 to 2019 February 27 during Sector 8 of the primary TESS mission. TESS observes over a wavelength range of 600 - 1100\,nm. \Nstar\ fell on TESS Camera 2/CCD 3 and data were reduced by the official SPOC pipeline \citep{SPOC}. We accessed the data through the MAST portal\footnote{\url{https://mast.stsci.edu/portal/Mashup/Clients/Mast/Portal.html}} and utilized the \texttt{PDCSAP\_FLUX}, which has had any spacecraft systematics removed \citep[see][for details]{SPOC}. The full TESS light curve of \Nstar\ spans $\sim$27.4\,days and contains four full transits. However for this work we consider only the sections of the light curve which coincide with the NGTS observations; $2458540.51961711 \leq BJD \leq 2458540.8531236$ for the transit night and $2458541.5199 \leq BJD \leq 2458541.84915$ for the non-transit night. This allows for a more direct comparison of the light curve and parameter precision available from the two data sets.

\begin{table}
	\centering
	\caption{Example table of the NGTS photometry of \Nstar. The full table is available online.}
	\label{tab:phot}
\begin{tabular}{cccc}
    BJD (TDB) &             Flux &  Flux Error & Cam\\
  (-2,450,000) &   &   &  \\ \hline
       8540.51964025 &  0.998963 & 0.008012  & 1\\ 
       8540.51979072 &  0.999378 & 0.007995  & 1\\ 
       8540.51994118 &  0.999461 & 0.007972  & 1\\ 
       8540.52009164 &  0.997650 & 0.007938  & 1\\ 
       8540.52024211 &  0.997247 & 0.007918  & 1\\ 
       8540.52039257 &  1.000107 & 0.007918  & 1\\ 
       8540.52054303 &  0.998085 & 0.007882  & 1\\ 
       8540.52068192 &  1.015229 & 0.007980  & 1\\ 
       8540.52083238 &  0.995757 & 0.007831  & 1\\ 
       8540.52098285 &  0.980196 & 0.007706  & 1\\ 

\hline
	\end{tabular}
\end{table}

\section{Analysis} \label{sec:analysis}
\subsection{NGTS Noise Properties} \label{sec:noise_props}
We used our NGTS observations of \Nstar\ on the night of 2019 February 26 to investigate noise correlation between NGTS telescopes. We studied three main properties of the noise of the individual NGTS light curves.

\subsubsection{Noise properties of the Individual Telescopes} \label{sec:rms_vs_nbins}
\begin{figure}
    \centering
    \includegraphics[width=\columnwidth]{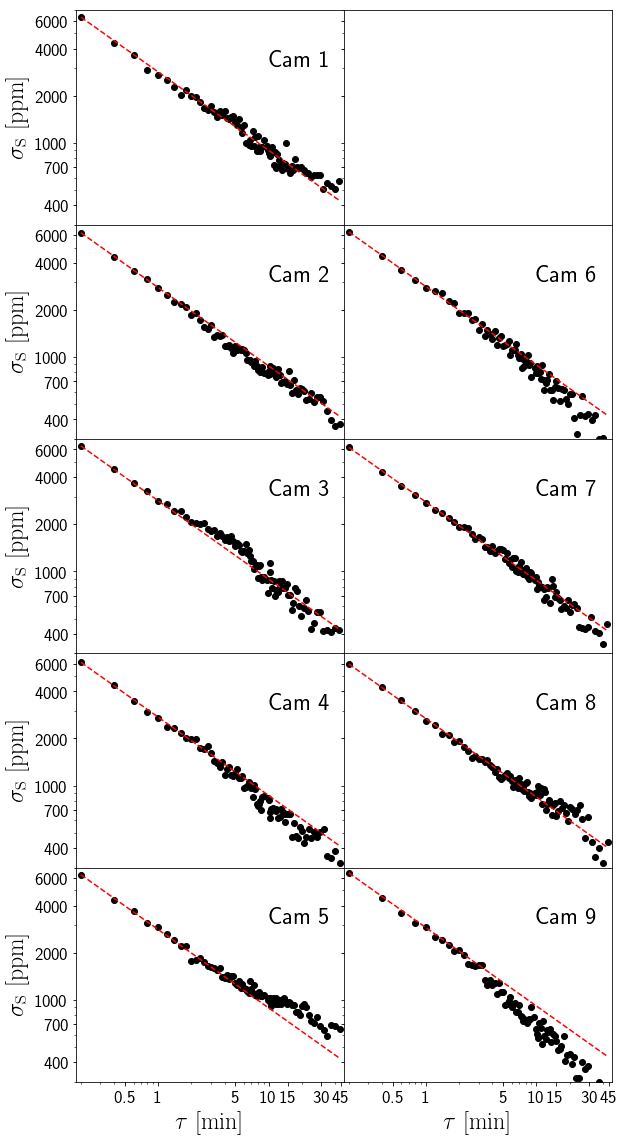}
    \caption{Variation of the light curve precision for each individual telescope light curve with the timescale over which the precision is calculated. The red dashed lines give the $1/\sqrt{\tau}$ Poisson noise scaling that would be expected for a pure white noise light curve and are scaled to the first point, which is a timescale of 13\,seconds - ie. unbinned data. Each panel gives the results from a single NGTS telescope, with the panel labels giving the ID of this telescope.}
    \label{fig:rmsvsbinsize}
\end{figure}

 To investigate the noise properties of the individual NGTS telescopes, we computed the RMS precision over a range of timescales, $\tau$, from unbinned cadence (13\,seconds) up to 45\,minutes.  We compare the RMS trend with timescale to that expected from purely white (Gaussian) noise.  Our results are plotted in Figure~\ref{fig:rmsvsbinsize}.  We find that the individual NGTS telescopes closely follow the predicted $1/\sqrt{\tau}$ white noise scaling.  This suggests that the noise in the NGTS light curves is dominated by Gaussian noise, and that we do not have any significant systematic noise on these timescales. In some telescopes we see deviations from the predicted Gaussian noise scaling, especially at larger $\tau$ values. This is a consequence of only having a small number of binned flux data points at large $\tau$ values.

This result gives us confidence that systematic noise is not a major hindrance to the effectiveness of this multi-telescope observing method. 

\subsubsection{Flux Correlations between Telescopes} \label{sec:flux_corr}
We performed a pair-wise correlation test of all the NGTS light curves of \Nstar\ to search for any obvious correlation in the photometric noise between the NGTS telescopes.  Since the NGTS telescopes all operate independently, the time-stamps for the nine individual light curves are not perfectly synchronous.  We therefore create a set of time-stamps at 13\,second intervals, between the start and end of the observations. Each telescope time series was then mapped to these new time-stamps before comparing to the other telescopes. This was done by assigning each flux data point to the new time-stamp which represents the smallest deviation in time from the actual mid-point of the exposure. This allows for a more accurate comparison between the fluxes from the pairs of telescopes.

\begin{figure*}
	\includegraphics[width=\textwidth]{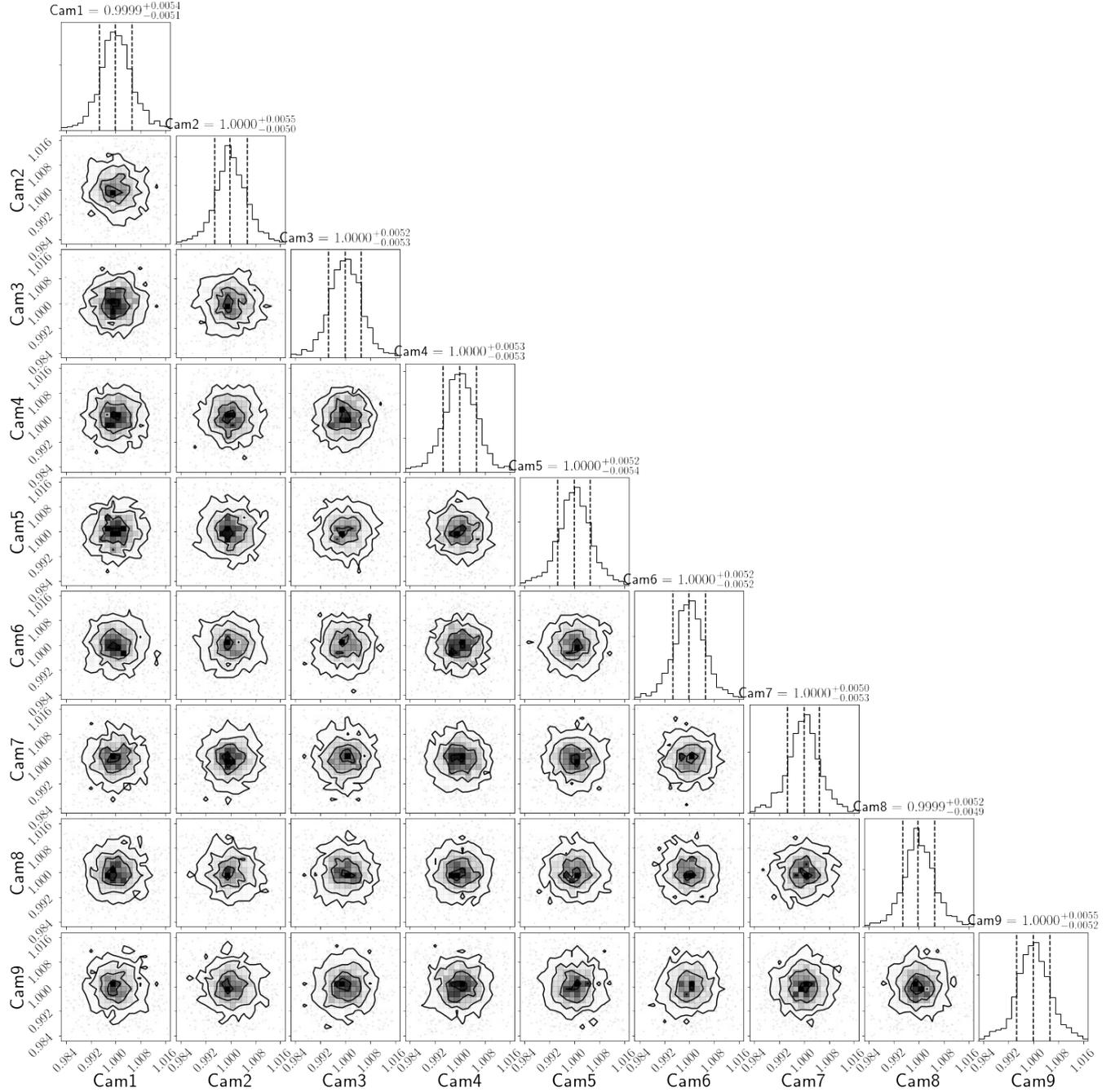}
    \caption{Corner plot displaying how the flux from \Nstar\ correlates for each possible pair of the nine NGTS telescopes. The headers give the median and 1$\sigma$ deviations of the flux from each telescope. The plot has been produced using \texttt{corner.py} \citep{corner}.}
    \label{fig:fluxcorner}
\end{figure*}

We find no flux correlation between any pair of telescopes - see Figure~\ref{fig:fluxcorner}.  We also note that the flux distribution for each telescope looks Gaussian. This again gives us confidence that the light curves are free from systematic noise.

\subsubsection{Combining Individual Telescope Data} 
Given individual NGTS light curves show uncorrelated Gaussian noise, by combining the light curves we expect a combined light curve with a RMS scatter, $\sigma_{m}$, given by:
\begin{equation}
\sigma_{m}=\sigma_{s}/\sqrt{N},
\label{eqn:sigma}
\end{equation} 
where $\sigma_{s}$ is the RMS scatter of a single telescope light curve, and $N$ is the number of telescopes combined.

\begin{figure}
    \centering
    \includegraphics[width=\columnwidth]{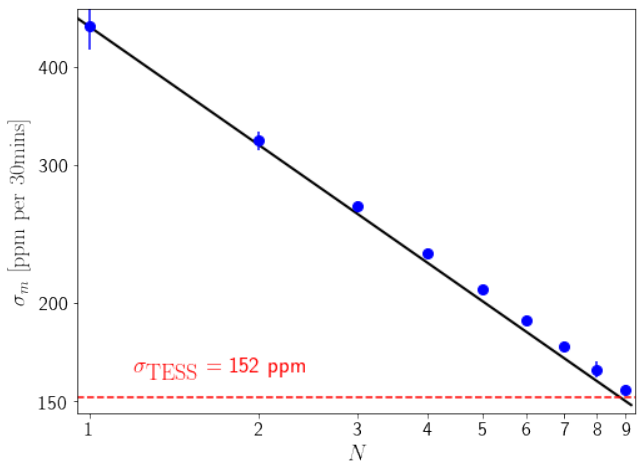}
    \caption{30\,minute RMS precision ($\sigma_{m}$) of the NGTS combined light curve for \Nstar\ as a function of the number of individual telescope light curves co-added. The blue points give the mean precision for each possible combination of $N$ individual light curves (see the text for details). The errorbars give the standard error on this mean. The solid black line shows the Gaussian $1/\sqrt{N}$ improvement in $\sigma_{m}$ for pure white noise, scaled from the $N=1$ data point.  The dashed red line shows the calculated RMS precision of the TESS light curve for \Nstar.}
    \label{fig:ncams_rms}
\end{figure}

For each value of $N$, we calculated $\sigma_{m}$ at a timescale of 30\,min for each possible telescope combination.  We then found the mean of the $\sigma_{m}$ for each value of $N$.  The results are set out in Figure~\ref{fig:ncams_rms}, showing that the calculated $\sigma_{m}$ values are very close to the uncorrelated Gaussian-noise expectation of $\sigma_{m}$ given by Equation~\ref{eqn:sigma}.  The deviation from the expected $\sigma_{m}$ for $N=9$ is just 5\,ppm.

\label{sec:rms_vs_ncams}

\subsubsection{Scintillation Noise}\label{sec:scin_noise}
As stated in Section~\ref{sec:intro}, the NGTS light curve noise for bright stars ($I \leq 10$) is expected to be dominated by scintillation noise \citep{NGTS_2018}. We tested this by analysing the variation in the flux RMS during the observation and comparing this variation to the predicted light curve noise. We calculated the predicted scintillation noise using the modified Young's approximation
\begin{equation}
    \centering
    \sigma^2_{Y} = 10 \times 10^{-6} \ C^2_Y \ D^{-4/3} \ t^{-1} \ (\sec z)^3 \ \exp(-2h_{obs}/H) \ ,
    \label{eqn:scin}
\end{equation}
where $D$ is the diameter of the telescope aperture (m), $t$ is the exposure time used (s), $z$ is the zenith distance, $h_{obs}$ is the altitude of the observatory (2440\,m for Paranal), and $H$ is the scale height of the atmospheric turbulence, which is taken to be 8000\,m \citep{osborn15}. $C_Y$\,(m$^{2/3}$s$^{1/2}$) is an empirical coefficient and $\sigma_Y$ is the dimensionless normalised scintillation noise \citep{young1967AJscintillation, osborn15}. 

At high airmass, differential refraction across the field-of-view will cause the comparison stars to move on the CCD relative to the target star. The slight differences in response of neighbouring pixels on the CCD will result in this movement causing an increase in the flux RMS at high airmass. Therefore, we expect the scintillation noise model (Equation~\ref{eqn:scin}) to not perfectly describe the observed light curve noise at high airmass. 

We modeled the noise in our data as a combination of scintillation noise and photon noise from both the target star and the sky background. We used only the data with airmass $\sec z < 1.2$ and fit for the coefficient $C_Y$, finding a value of $C_Y \ = \ 1.57\pm0.06$. This is in good agreement with the results from \citet{osborn15}. They find a median value of $C_Y = 1.56$ for Paranal, with 1st and 3rd quartiles of 1.27 and 1.90. 

\begin{figure}
    \centering
    \includegraphics[width=\columnwidth]{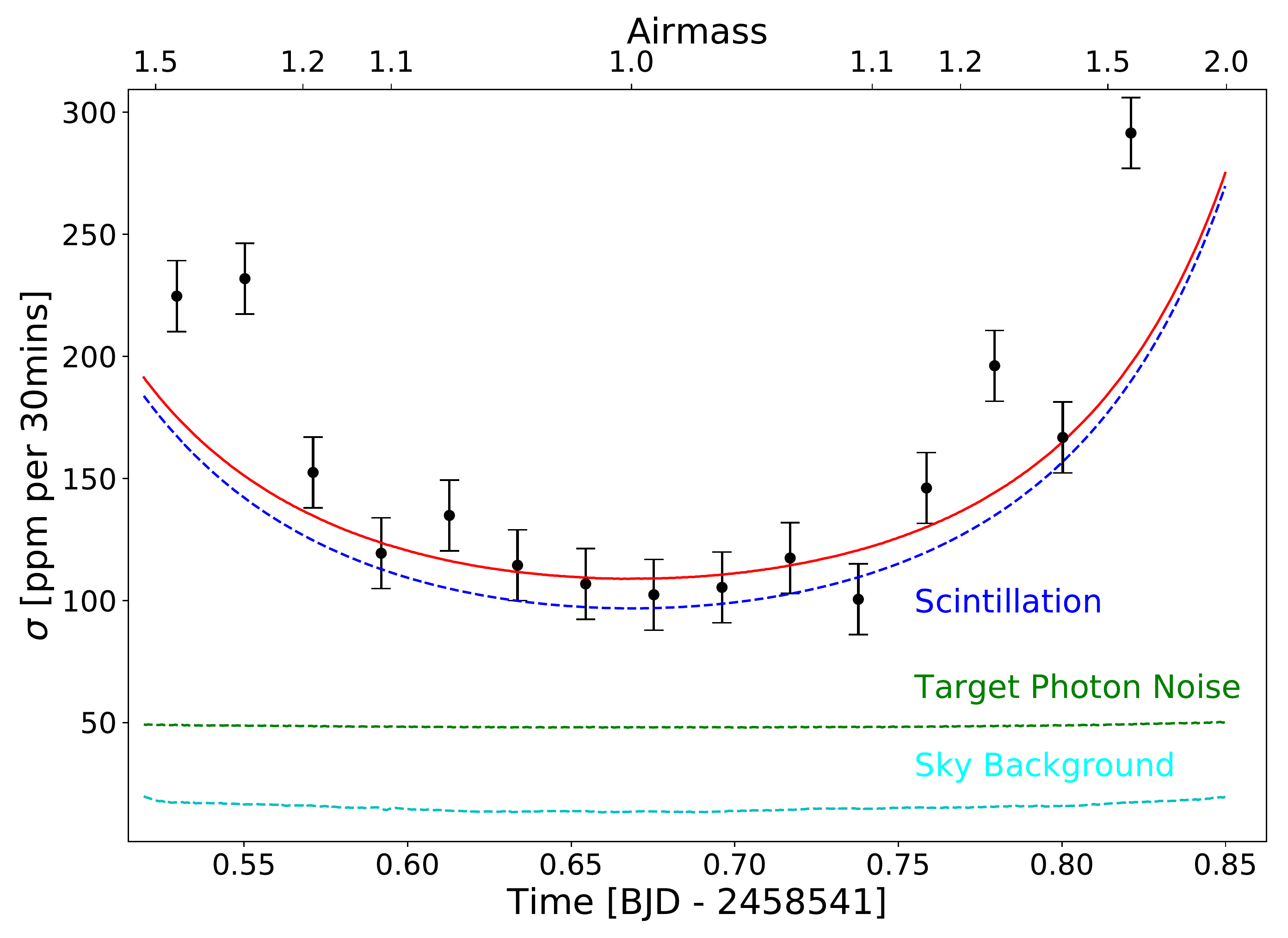}
    \caption{Variation of the flux RMS for the combined NGTS light curve. The total noise model (solid red) includes the scintillation noise (blue dashed) and the photon noise from the target star (green dashed) and the sky background (cyan dashed).}
    \label{fig:noise_models}
\end{figure}

The flux RMS and the noise model are displayed in Figure~\ref{fig:noise_models}, and we see that the noise in the NGTS light curve is dominated by the scintillation noise, as predicted. The variation in this noise during the night is well described by Young's approximation for low airmass. The predicted deviation of the flux RMS values from the noise model at high airmass can also be clearly seen.

\subsection{Fitting Individual NGTS Light Curves} \label{sec:NGTSfitind}
In order to determine the transit parameters of \Nplanet\ from our NGTS data from the UT night of 2019 February 25, we simultaneously detrended the individual NGTS telescope light curves with respect to external parameters and fitted each transit using \texttt{batman} \citep{BATMAN}.  We tested a number of external parameters for detrending, including airmass, target pixel position, sky background, and time.  We found that only detrending with respect to time significantly improved the log likelihood, $\ln\,z$, of the final model. 

For the \texttt{batman} transit model, we used the following free parameters: the time of the transit centre, $T_C$, the orbital period, $P$, the planet-to-star radius ratio, $R_p/R_*$, the impact parameter, $b$, and the stellar density, $\rho_*$. We used a quadratic limb-darkening law, and fitted the parametrized coefficients $q_1$ and $q_2$ from \citet{KippingLD2013}. For these coefficients, we used uniform priors between 0 and 1, in order to ensure physically realistic limb-darkening profiles \citep{KippingLD2013}.

As the NGTS data cover just a single transit, they alone do not place any constraints on the period. However, \citet{WASP-166} were able to very tightly constrain the period due to the very long time baseline afforded to them by the combination of WASP and TESS photometry, as well as their precise measurement of the Rossiter-McLaughlin effect for \Nstar. Therefore we place a $1\sigma$ Gaussian prior on the period, based on the period and uncertainty reported in \citet{WASP-166}. We also utilise the stellar parameters from version 8 of the TIC \citep{TESS_CAT} to place a Gaussian prior on $\rho_*$ with a mean and standard deviation of $0.58 \pm 0.11 \rho_{\odot}$. In addition, we use a uniform prior to ensure that $b \geq 0$. For the remaining parameters, we simply impose further uniform priors to ensure that they take physically realistic values. \citet{WASP-166} find a 2$\sigma$ upper limit of the eccentricity of $e < 0.07$, and so they adopt a circular orbit. Based on this, and the fact that our photometric data provide little information on the eccentricity of the orbit, we also adopt $e = 0$. The modeling was performed using an MCMC sampling method implemented using the \textit{emcee} Ensemble Sampler \citep{emcee}. We ran 50 walkers for 15000 steps as a burn-in process, and then a further 5000 draws were made to sample the posterior for each chain. The chains were inspected and found to be well mixed. We plot the best fit models for each individual light curve in Figure~\ref{fig:ind_lcs}.

\begin{figure}
    \centering
    \includegraphics[width=\columnwidth]{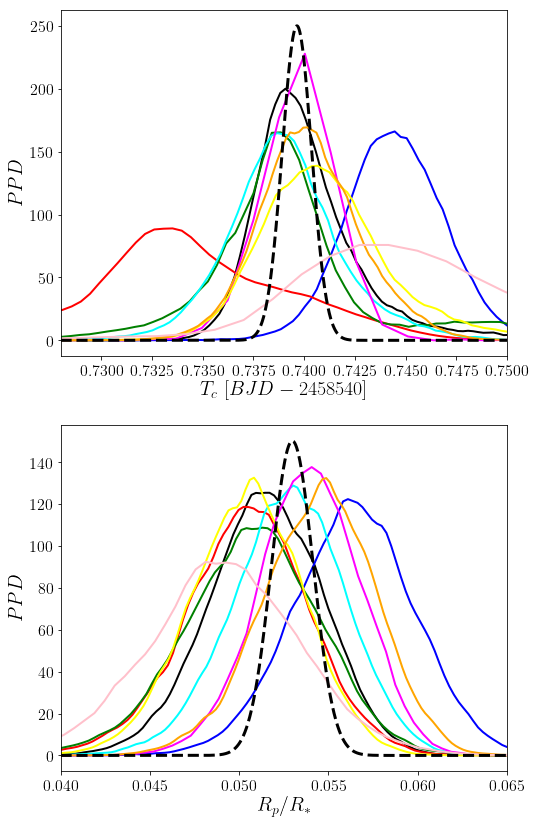}
    \caption{Posterior Parameter Distributions (PPDs) for the two main free parameters included in the fits: $T_C$ (top panel) and $R_p/R_*$ (bottom panel). The solid lines give the PPDs for the individual telescopes, with each colour corresponding to the same telescope in each panel. These PPDs are normalised such that the area enclosed is equal to 1. The dashed black curves give the PPDs from the fit to the full combined data set. These curves are scaled such that the area enclosed is less than 1 for clarity.}
    \label{fig:post_comp}
\end{figure}

In analysing these fits, we focus on the two main parameters which can be obtained from a single transit: the planet-to-star radius ratio, $R_p/R_*$, and the time of the transit centre, $T_c$. In Figure~\ref{fig:post_comp}, we compare the posterior parameter distributions (PPDs) for these two parameters from the modelling of the nine individual NGTS light curves.

In terms of the individual PPDs, we see good agreement between the obtained parameter values for the two parameters. For $R_p/R_*$, the PPDs behave as might be expected, with the single telescope PPDs being scattered around the "true" value. The average single telescope uncertainty in $R_p/R_*$ is $\pm\,0.0034$. The weighted mean of the nine $R_p/R_*$ measurements is $0.05245\,\pm\,0.00110$.

For $T_C$, we again see the expected distribution of single light curve values, but with a couple of outliers. These outliers are present for $T_C$ but not $R_p/R_*$ since the measured value of $T_C$ is dependant upon the measured ingress and egress times. These sections of the light curve are each only 13\,minutes long for a transit of \Nplanet, and so a few out-lying points in these sections of the light curve strongly bias the measured $T_C$ value. On the other hand, as the section of the light curve from which $R_p/R_*$ is measured is $\sim$3.5\,hours long, the same number of out-lying points will not have a noticeable effect on the measured $R_p/R_*$ value. The average single telescope uncertainty from all nine measurements of $T_C$ is $\pm\,0.00318$\,days. We find a weighted mean of the nine measurements of $2458540.74035\,\pm\,0.00088$.

\begin{figure}
    \centering
    \includegraphics[width=\columnwidth]{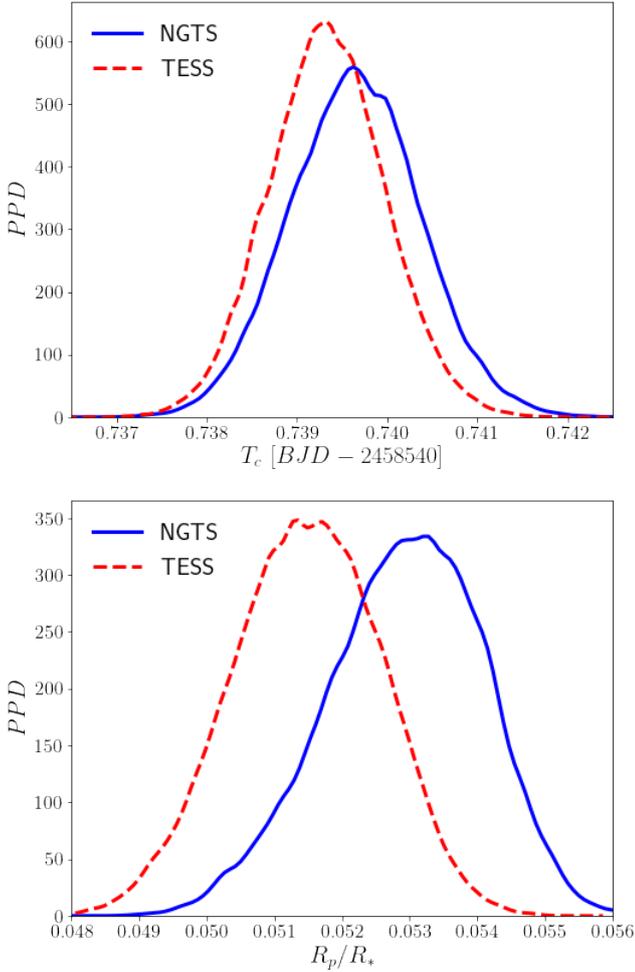}
    \caption{Same as Fig.~\ref{fig:post_comp} but comparing the PPDs from the fit to the combined NGTS light curve (solid, blue online) and the fit to the TESS transit on 2020 February 25 (dashed, red online)}.
    \label{fig:post_ngts_vs_tess}
\end{figure}

\subsection{Fitting Combined Light Curve} \label{sec:NGTScombined}

\begin{figure*}
	\includegraphics[width=\textwidth]{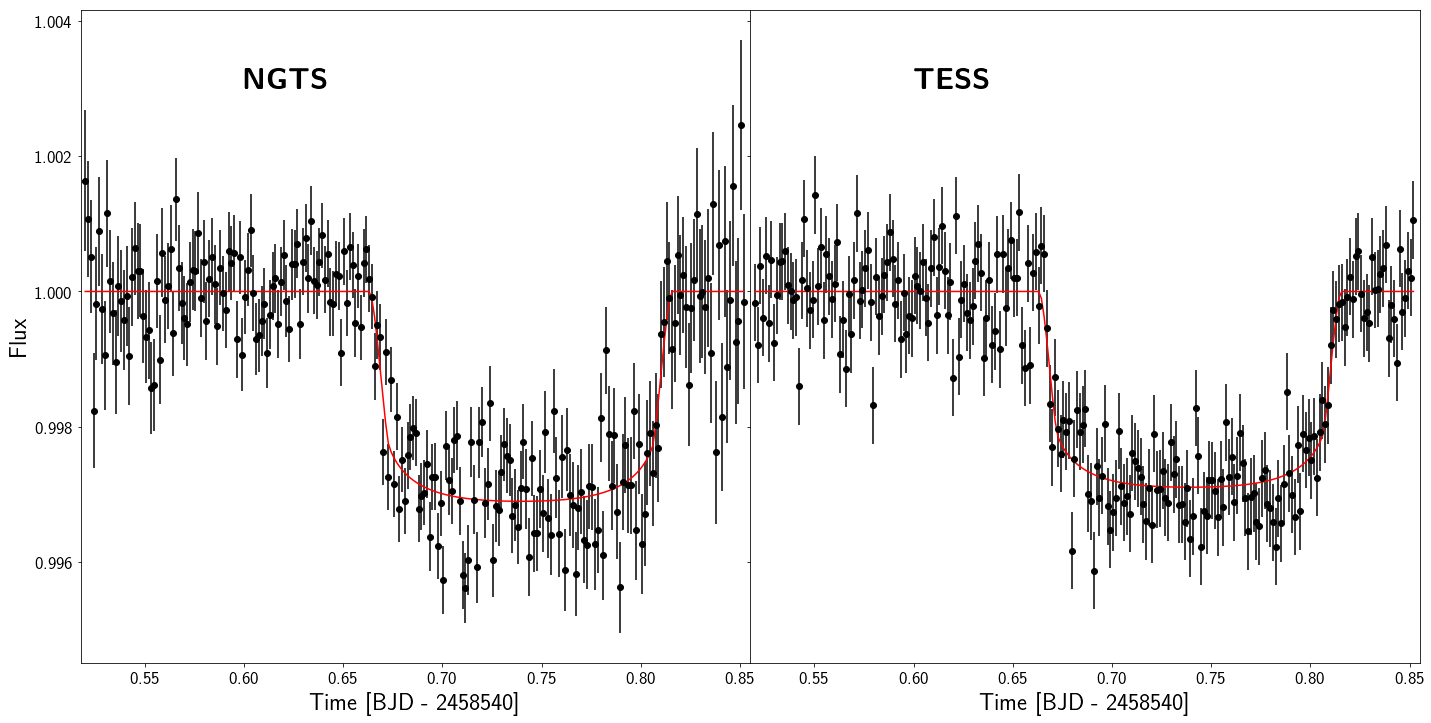}
    \caption{Light curves showing the 2019 February 25 transit of \Nplanet. \textbf{Left:} combined light curve obtained with a simultaneous observation using nine NGTS telescopes. The black circles show the data binned to 2\,minutes, for visual comparison with the 2\,minute cadence TESS data. The solid red line gives the best fit transit model from the modelling process (section \ref{sec:NGTScombined}). \textbf{Right:} Black circles show the unbinned 2\,minute cadence TESS data. The solid red line again gives the transit model from the modelling process (section \ref{sec:TESSanalysis}).}
    \label{fig:LC}
\end{figure*}

In addition to fitting each of the individual light curves, we also fitted the combined light curve.  For this fitting we also included a constant offset, $c$, to account for the position of the out-of-transit flux baseline.

The combined NGTS light curve and the resultant transit model from this modelling process are displayed in Figure~\ref{fig:LC}. The system parameters obtained are given in Table \ref{tab:planetary}, where the parameter values reported are the median values from the posterior distributions, with the corresponding $1\sigma$ uncertainties.

We compare this combined light curve fit to the individual light curves fits from Section~\ref{sec:NGTSfitind}. The PPD of the combined light curve fit is plotted in Figure~\ref{fig:post_comp}.  The average single telescope uncertainty in $R_p/R_*$ is $\pm\,0.0034$, while the uncertainty in $R_p/R_*$ from the combined light curve is $\pm\,0.0012$. Thus we see that the uncertainty in the measured value of $R_p/R_*$ from the combined light curve is reduced by a third, as expected from the $1/\sqrt{N}$ scaling. In addition, the combined light curve uncertainty is comparable to the error on the weighted mean of the nine individual light curve measurements. We also find that the $R_p/R_*$ value derived from the combined light curve is consistent with the nine light curve weighted mean at the $1\sigma$ level.

The uncertainty in $T_C$ measured from the combined light curve is $\pm\,0.0007$\,days, which is a reduction from the average of the individual light curve uncertainties found in Section~\ref{sec:NGTSfitind} ($\pm\,0.00318$\,days) of more than a third. The two individual light curves with the high $T_C$ uncertainties inflate the average single telescope uncertainty. However, the fit to the combined light curve is less affected by the out-lying points which result in the higher single telescope uncertainties.

As with $R_p/R_*$, the $T_C$ value derived from the combined light curve is consistent with the nine light curve weighted mean at the $1\sigma$ level. We note that the error on this weighted average is slightly higher than the combined light curve $T_C$ uncertainty. This is again likely a result of the out-lying individual $T_C$ measurements having more of an effect on the weighted mean than is had by the out-lying data points in these light curves on the combined light curve $T_C$ measurement.

\subsection{Fitting TESS Data} \label{sec:TESSanalysis}
In order to make a direct comparison to the results of modelling the NGTS data, we model the TESS data with a process identical to that for the combined NGTS light curve set in Section~\ref{sec:NGTScombined}.  We select just the portion of the TESS data obtained simultaneously with our NGTS data ($2458540.51961711 \leq BJD \leq 2458540.8531236$).  The TESS data and the resulting best fit transit model are displayed in Figure~\ref{fig:LC}, with the parameter values given in Table~\ref{tab:planetary}. 

\subsection{NGTS and TESS Comparison}\label{sec:ng_tess_comp}
The main parameters of interest for this comparison are $T_C$ and $R_p/R_*$, as they both can be constrained by a single transit. We plot the PPDs for these parameters in Figure~\ref{fig:post_ngts_vs_tess}. 

We find the two values of $T_C$ from NGTS and TESS to be consistent to within $0.25\sigma$ of each other.  The value of $R_p/R_*$ measured from the TESS transit is found to be a lower than the value measured from the NGTS data, with a difference of $0.9\sigma$. The $R_P/R_*$ values from NGTS and TESS are formally consistent to $1\sigma$, but we note that a similar modelling of the first transit of \Nplanet\ in the TESS light curve yields a value of $R_p/R_* = 0.05235\pm0.00105$. This value represents a difference from the NGTS value of just $0.4\sigma$. We fitted the limb darkening coefficients during the modelling, therefore any differences in measured $R_p/R_*$ values will not be due to any depth difference caused by the differences between the TESS and NGTS pass bands. \citet{WASP-166} measure a value of $R_p/R_* = 0.0530\pm\,0.0007$, which is consistent with the value measured from the NGTS transit and the first TESS transit, but is discrepant with the value from the TESS transit on 2020 February 25.

During Sector 8 of the TESS mission, a few days prior to the 2020 February 25 transit, an instrument anomaly caused the heaters to switch on\footnote{\url{https://archive.stsci.edu/missions/tess/doc/tess_drn/tess_sector_08_drn10_v02.pdf}}. The resulting temperature increase affected both the camera focal plane scale and the individual CCD mean black levels. It is probable that this resulted in a problem with the systematic error corrections in SPOC pipeline, which could have induced the slightly shallower transit depth observed for this transit. 

\section{Discussion} \label{sec:discussion}
The NGTS photometric noise for bright stars is dominated by atmospheric scintillation \citep{NGTS_2018}.  Scintillation behaves as white noise on the timescales of exoplanet transits \citep{fohring19}, and indeed the NGTS light curves in this study confirm this: see Section~\ref{sec:noise_props}.  We also find that there is no correlation between the photometric noise from individual NGTS telescopes - see Figs.~\ref{fig:fluxcorner} \& \ref{fig:ncams_rms}.  This indicates that the spacing between the telescopes (approximately 2\,m) is enough to ensure that the light paths through the atmosphere are separated enough to ensure that they do not result in correlated scintillation noise. The fact that the NGTS telescopes are well spaced is critical for the success of these observations.

For these bright stars, the noise will be dominated by scintillation and photon noise, which will both result in the precision increasing by $\sqrt{N}$ when combining telescopes.  However, from \citet{NGTS_2018}, CCD readout and sky background become significant noise sources for $I>13$.  This suggests that combining NGTS telescopes would be less advantageous for fainter targets. 

We find that we can achieve a precision for $T_C$ on a single transit of 1\,minute for a transit of this bright star. This will allow NGTS multi-telescope observations to measure $T_C$ with a precision of a few minutes for systems of multiple rocky planets, orbiting bright host stars. The TESS mission has already revealed many candidate systems of this type, such as TOI-175 \citep{toi175Kostov, toi175Cloutier}, TOI-178 \citep{toi178}, and TOI-270 \citep{toi270Guenther}. These stars are 2 to 3 magnitudes fainter than \Nstar, and so we expect lower photometric precision than we achieved for \Nstar. A lower photometric precision is likely to result in a higher uncertainty on the transit timing \citep{carter08}.

We also find in Section~\ref{sec:NGTScombined} that the measured value of $T_C$ from a single telescope light curve can be biased by out-lying data points during ingress or egress. This in turn affects the $T_C$ value that is derived from a weighted mean of the individual $T_C$ measurements. By modelling all the data together as a combined light curve, we find that the measured $T_C$ value and uncertainty is more resistant to these flux outliers.

Observing simultaneously with multiple telescopes also grants some protection against technical faults with individual telescope systems. This is especially of importance for high priority candidates with not many opportunities for observations, such as long period candidates (eg. TOI-222, \citet{Lendl2020}; TIC-238855958, \citet{gill20}).

The TESS mission has recently been extended for two more years\footnote{\url{https://heasarc.gsfc.nasa.gov/docs/tess/announcement-of-the-tess-extended-mission.html}}. As such, the Southern sky will be re-observed by TESS between July 2020 and June 2021. TESS observes in the anti-Sun direction, and so this extended mission will provide many more opportunities for simultaneous ground and space observations for bright planet hosting stars.

\section{Conclusion}\label{sec:conclusion}
Our data have shown that the noise in NGTS bright star light curves is dominated by scintillation noise. We have shown that this noise is Gaussian, and is uncorrelated between the telescopes. This allows us to combine simultaneous observations with multiple telescopes to obtain ultra-high precision light curves of individual exoplanet transits. We can combine this technique with the wide field-of-view of the NGTS cameras to achieve some of the highest precisions from the ground for the brightest stars.

We can use this technique to achieve transit timing on order of minutes for planets orbiting bright stars, with transit depths on the order $\sim1000$s\,ppm. As a result, this technique will enable NGTS to measure any significant transit timing variations in multi-planet systems with similarly bright host stars. The precise transit ephemerides achievable with this observing method will also allow NGTS to monitor the transit timing variations of short period Jupiter-like planets to search for signs of orbital decay \citep[eg.][]{baluev19tiddecay, yee20wasp12, patra20tidaldecay}. In addition to measuring transit timing variations, the precise ephemerides achievable with NGTS multi-telescope observations will also be of use for the scheduling of future transmission spectroscopic measurements, and other characterisation efforts.

Over the next few years, hundreds of rocky planet candidates orbiting bright stars will be detected by TESS. The confirmation of these planets and the measurement of their masses will contribute to the TESS Level 1 Science goal of measuring the mass of 50 planets with a radius $\leq 4\,R_{\oplus}$. We have demonstrated that NGTS can achieve the same precision as space-based photometry from the ground through simultaneous multi-telescope observations. As such, NGTS will contribute significantly to the confirmation of these and other rocky planets around bright stars, both by confirming the transit signal and by measuring the mass by detecting any TTV signals.

\begin{table*}
	\centering
	\caption{Planetary System properties for \Nstar}
	\begin{tabular}{lcc} % six columns, alignment for each
	\hline
	\multicolumn{3}{l}{}\\
	\textbf{Planetary Parameters}	&	\textbf{Value (NGTS)}    &    \textbf{Value (TESS)} \\
	\hline
    \multicolumn{3}{l}{\textbf{Fitted Parameters}}\\
    \hline
    Tc (BJD)		&	\NtcNGTS	&	\NtcTESS	\\
	Orbital Period (days)			&	\NperiodNGTS    &   \NperiodTESS	\\
    $R_p / R_{\ast}$ & \NrratioNGTS & \NrratioTESS \\
	b & \NimpactNGTS & \NimpactTESS \\
	\rhostar (\rhosun) & \NstardensityNGTS & \NstardensityTESS \\
	$q_1$ & \NqoneNGTS & \NqoneTESS \\
	$q_2$ & \NqtwoNGTS & \NqtwoTESS \\
    \\
    \hline
    \multicolumn{3}{l}{\textbf{Derived Parameters}}\\
    \hline
    a (\rstar)     &  \NaoverrNGTS & \NaoverrTESS \\
    a (AU)      & \NauNGTS & \NauTESS \\
    inc ($^{\circ}$) & \NincNGTS & \NincTESS \\
    \\
    \hline
    \multicolumn{3}{l}{\textbf{Fixed Parameters}}\\
    \hline
    e & 0. & 0. \\
    $\omega$ ($^{\circ}$) & 90. & 90. \\
	\hline
    \end{tabular}
    \label{tab:planetary}
\end{table*}

\section*{Acknowledgements}

Based on data collected under the NGTS project at the ESO La Silla Paranal Observatory.  The NGTS facility is operated by the consortium institutes with support from the UK Science and Technology Facilities Council (STFC)  project ST/M001962/1. 
This paper includes data collected by the TESS mission. Funding for the TESS mission is provided by the NASA Explorer Program.
JSJ is supported by funding from Fondecyt through grant 1161218 and partial support from CATA-Basal (PB06, Conicyt).
This work has been in particular carried out in the frame of the National Centre for Competence in Research "PlanetS" supported by Swiss National Science Foundation.
This work has made use of data from the European Space Agency (ESA) mission
{\it Gaia} (\url{https://www.cosmos.esa.int/gaia}), processed by the {\it Gaia}
Data Processing and Analysis Consortium (DPAC,
\url{https://www.cosmos.esa.int/web/gaia/dpac/consortium}). Funding for the DPAC
has been provided by national institutions, in particular the institutions
participating in the {\it Gaia} Multilateral Agreement.
This publication makes use of data products from the Two Micron All Sky Survey, which is a joint project of the University of Massachusetts and the Infrared Processing and Analysis Center/California Institute of Technology, funded by the National Aeronautics and Space Administration and the National Science Foundation.
%\\
%%%%%%%%%%%%%%%%%%%%%%%%%%%%%%%%%%%%%%%%%%%%%%%%%%



\bibliographystyle{mnras}
\bibliography{wasp166.bib} % if your bibtex file is called example.bib

% \begin{thebibliography}{99}
% \bibitem[\protect\citeauthoryear{Author}{2012}]{Author2012}
% Author A.~N., 2013, Journal of Improbable Astronomy, 1, 1
% \bibitem[\protect\citeauthoryear{Others}{2013}]{Others2013}
% Others S., 2012, Journal of Interesting Stuff, 17, 198
% \end{thebibliography}

%%%%%%%%%%%%%%%%%%%%%%%%%%%%%%%%%%%%%%%%%%%%%%%%%%

%%%%%%%%%%%%%%%%% APPENDICES %%%%%%%%%%%%%%%%%%%%%

% \appendix

% \section{Some extra material}

% If you want to present additional material which would interrupt the flow of the main paper,
% it can be placed in an Appendix which appears after the list of references.

%%%%%%%%%%%%%%%%%%%%%%%%%%%%%%%%%%%%%%%%%%%%%%%%%%

% Don't change these lines
\bsp	% typesetting comment
\label{lastpage}
\end{document}

%% file: commands.tex
\newcommand{\masy}{mas\,y$^{-1}$}

\newcommand{\mstar}{\mbox{$M_{\ast}$}}
\newcommand{\rstar}{\mbox{$R_{\ast}$}}
\newcommand{\rhostar}{\mbox{$\rho_{\ast}$}}

\newcommand{\rjup}{\mbox{$R_{J}$}}
\newcommand{\msun}{\mbox{$M_{\odot}$}}
\newcommand{\rsun}{\mbox{$R_{\odot}$}}
\newcommand{\rhosun}{\mbox{$\rho_{\odot}$}}

%% file: properties.tex
\newcommand{\Nstar}{WASP-166}
\newcommand{\NTICID}{408310006}
\newcommand{\NTOIID}{576}
\newcommand{\NRA}{\mbox{$09^{\rmn{h}} 39^{\rmn{m}} 30\fs09$}} %2MASS
\newcommand{\NDec}{\mbox{$-20\degr 58\arcmin 56\farcs 9$}} %2MASS
\newcommand{\NpropRA}{\mbox{$-55.082\pm0.072 $}} %UCAC4 
\newcommand{\NpropDec}{\mbox{$10.927\pm0.069$}} %UCAC4
\newcommand{\Nplx}{\mbox{$8.7301 \pm 0.0448$}}

\newcommand{\NstarradiusTIC}{\mbox{$1.25\pm0.06$}}
\newcommand{\NstarmassTIC}{\mbox{$1.14\pm0.15$}}

\newcommand{\NstardensityNGTS}{\mbox{$0.593\,^{+0.082}_{-0.080}$}}
\newcommand{\NstardensityTESS}{\mbox{$0.627\,^{+0.074}_{-0.081}$}}
\newcommand{\NVmag}{$9.351 \pm 0.002$}
\newcommand{\NBmag}{$9.940 \pm 0.034$}

\newcommand{\NGAIAmag}{$9.257 \pm 0.00031$}

\newcommand{\NJmag}{$8.350 \pm 0.021$}
\newcommand{\NHmag}{$8.135 \pm 0.033$}
\newcommand{\NKmag}{$8.032 \pm 0.023$}
\newcommand{\NWmag}{$7.999 \pm 0.025$}
\newcommand{\NWWmag}{$8.054 \pm 0.019$}
\newcommand{\NTESSmag}{$8.8685 \pm 0.0062$}
\newcommand{\NshortTmag}{$8.87$}

\newcommand{\Nplanet}{WASP-166\,b}
\newcommand{\NperiodNGTS}{\mbox{$5.4435399 \pm 0.0000028$}}
\newcommand{\Nperiodshort}{\mbox{$5.4435$}}
\newcommand{\NtcNGTS}{\mbox{$2458540.739646\,^{+0.000715}_{-0.000726}$}}
\newcommand{\NradiusNGTS}{\mbox{$0.64\pm0.03$}}%
\newcommand{\NrratioNGTS}{\mbox{$0.05298\,^{+0.00110}_{-0.00123}$}}
\newcommand{\NauNGTS}{\mbox{$0.064\pm 0.003$}}%
\newcommand{\NaoverrNGTS}{\mbox{$11.00\,^{+0.47}_{-0.51}$}}%
\newcommand{\NincNGTS}{\mbox{$87.80\,^{+0.67}_{-0.59}$}}

\newcommand{\NimpactNGTS}{\mbox{$0.423\,^{+0.090}_{-0.118}$}}%
\newcommand{\NqoneNGTS}{\mbox{$0.26\,^{+0.31}_{-0.15}$}}
\newcommand{\NqtwoNGTS}{\mbox{$0.16\,^{+0.31}_{-0.12}$}}
\newcommand{\NtcTESS}{\mbox{$2458540.739389\,^{+0.000721}_{-0.000670}$}}
\newcommand{\NperiodTESS}{\mbox{$5.4435402\,^{+0.0000027}_{-0.0000030}$}}
\newcommand{\NrratioTESS}{\mbox{$0.0515\,^{+0.0011}_{-0.0011}$}}
\newcommand{\NimpactTESS}{\mbox{$0.398\,^{+0.093}_{-0.111}$}}
\newcommand{\NauTESS}{\mbox{$0.065 \pm 0.003$}}
\newcommand{\NaoverrTESS}{\mbox{$11.14\,^{+0.42}_{-0.50}$}}
\newcommand{\NincTESS}{\mbox{$87.95\,^{+0.62}_{-0.59}$}}
\newcommand{\NqoneTESS}{\mbox{$0.33\,^{+0.30}_{-0.17}$}}
\newcommand{\NqtwoTESS}{\mbox{$0.16\,^{+0.28}_{-0.12}$}}